\documentclass[entropy,article,accept,moreauthors,pdftex]{Definitions/mdpi} 

\firstpage{1} 
\pubvolume{xx}
\issuenum{1}
\articlenumber{5}
\pubyear{2020}
\copyrightyear{2020}
\history{}

\usepackage{amsfonts}
\usepackage[T1]{fontenc}
\usepackage{graphicx}
\usepackage{amsmath}
\usepackage{float} 
\usepackage{amssymb}
\usepackage{color}
\usepackage[version=3]{mhchem}
\usepackage{url}

\usepackage{multirow}

\usepackage{placeins}

\hypersetup{
colorlinks=true,
linkcolor=blue,         
citecolor=blue,         
filecolor=magenta,      
urlcolor=red
}

\Title{Dynamics of Ultracold Bosons in Artificial Gauge Fields: Angular Momentum, Fragmentation, and the Variance of Entropy}

\Author{Axel U.~J. Lode\orcidA$^1$, Sunayana Dutta\orcidB$^{2,3}$, Camille L\'{e}v\^{e}que\orcidC$^{4,5}$}

\address{$^{1}$ \quad Physikalisches Institut, Albert-Ludwigs-Universit\"{a}t  Freiburg, Hermann-Herder-Stra{\ss}e 3, D-79104  Freiburg, Germany;\\ 
$^2$ \quad Department of Mathematics, University of Haifa, Haifa 3498838, Israel;\\	
$^3$ \quad Haifa Research Center for Theoretical Physics and Astrophysics,
University of Haifa, Haifa 3498838, Israel;\\
$^4$ \quad Vienna Center for Quantum Science and Technology,
Atominstitut, TU Wien, Stadionallee 2, 1020 Vienna, Austria;\\
$^5$ \quad Wolfgang Pauli Institute c/o Faculty of Mathematics,
University of Vienna, Oskar-Morgenstern Platz 1, 1090 Vienna, Austria
}

\corres{Correspondence: auj.lode@gmail.com (A.U.J.L.)}

\abstract{We consider the dynamics of two-dimensional interacting ultracold bosons triggered by suddenly switching on an artificial gauge field. The system is initialized in the ground state of a harmonic trapping potential. As a function of the strength of the applied artificial gauge, we analyze the emergent dynamics by monitoring the angular momentum, the fragmentation as well the entropy and variance of the entropy of absorption or single-shot images. We solve the underlying time-dependent many-boson Schrödinger equation using the multiconfigurational time-dependent Hartree method for indistinguishable particles (MCTDH-X).
We find that the artificial gauge field implants angular momentum in the system. Fragmentation -- multiple macroscopic eigenvalues of the reduced one-body density matrix -- emerges in sync with the dynamics of angular momentum: the bosons in the many-body state develop non-trivial correlations. Fragmentation and angular momentum are experimentally difficult to assess; here, we demonstrate that they can be probed by statistically analyzing the variance of the image entropy of single-shot images that are the standard projective measurement of the state of ultracold atomic systems.}


\begin{document}

\section{Introduction}
Since the first realization of Bose-Einstein condensates in 1995~\cite{davis:95,anderson:95,bradley:95}, ultracold atoms have become a standard probe for analog quantum simulations: due to the tunability and flexibility of these quantum states of matter, they can be manipulated to behave like other systems, for instance, condensed matter systems which are not as flexible or easy to observe. Popular examples include the realization of the quantum simulation of the superfluid-to-Mott-insulator transition~\cite{jaksch:98,greiner:02}, quantized conductance~\cite{krinner:15,corman:19}, the Dicke model~\cite{baumann:10,lode:17}, and magnetism realized via artificial gauge fields for ultracold atoms~\cite{dalibard.rmp:11}. 

Such artificial gauge fields can make the neutral ultracold atoms behave as if they were charged particles experiencing a magnetic field and were investigated experimentally and theoretically with an external lattice potential~\cite{aidelsburger.thesis:16,aidelsburger:11,lim:08} or without one~\cite{spielman:09,lin:09,lin.prl:09}.

In this paper, we investigate the physics of a two-dimensional system of harmonically trapped interacting ultracold bosons quenched with an artificial magnetic field (AMF) from a \emph{many-body} point of view.
The time-dependent Gross-Pitaevskii mean-field theory~\cite{gross:61,pitaevskii:61} is the most widespread tool to theoretically model many-body systems of ultracold bosonic atoms subject to an AMF. This approach recovers many of the physical phenomena observed, but neglects correlations by its construction using a mean-field ansatz; here, we go beyond mean-field and use the multiconfigurational time-dependent Hartree method for bosons (MCTDH-B)~\cite{streltsov:07,alon:08,alon.jcp:07} to approximate the solution of the Schrödinger equation for ultracold atoms subject to an AMF. MCTDH-B is a method from the MCTDH family methods~\cite{beck:00,wang:03,manthe:08,wang:15,manthe:17,manthe:17_review} for indistinguishable particles (MCTDH-X)~\cite{zanghellini:03,schmelcher:13,schmelcher:17,beck:00,haxton.pra2:15,alon:07_mix,alon:12,miyagi:13,miyagi:17,leveque:17,leveque:18,lode:20} that is able to self-consistently describe correlation effects like quantum correlations~\cite{lode:14,klaiman:15,klaiman:16,klaiman:16b,klaiman:18,alon:18b,alon:19,tsatsos:17,lode:17,roy:18,lode.njp:18,chatterjee:18,chatterjee:19,bera:19,lin:19,lin:20,lin.prar:20,chatterjee:20}. A key focus of the applications of MCTDH-B has been the emergence of fragmentation~\cite{nozieres:82,spekkens:99,mueller:06}, where the reduced one-body density matrix has multiple significant eigenvalues, see, for instance, Refs.~\cite{streltsov.jpb:09,streltsov.pra:09,streltsov.prl:08,sakmann:11,streltsov.prl:11,lode2:12,beinke:15,weiner:17,dutta:19}.
To obtain the results presented in this work, we used the MCTDH-X software hosted at \url{http://ultracold.org}, see Refs.~\cite{fasshauer:16,lode2:16,lode3:16,lin:20,lode:20,ultracold}.

Our paper is structured as follows: in Sec.~\ref{sec:HAM} we introduce the Hamiltonian and the MCTDH-X method we use, in Sec.~\ref{sec:QOI} we discuss the observables that we are using in Sec.~\ref{sec:RES} to investigate the dynamics of ultracold atoms in an AMF; Sec.~\ref{sec:END} summarizes our conclusions and provides an outlook.

\section{Hamiltonian and Methods}\label{sec:HAM}

We consider a system of bosonic particles with two-body interactions in two spatial dimensions. The state of the bosons is initialized in the ground state of a parabolic trap without an AMF present. Subsequently, the system is quenched by turning on suddenly an artificial gauge field corresponding to a homogeneous AMF perpendicular to the plane in which the bosons are trapped.

For the sake of clarity of presentation, we will omit the dependence of quantities on time $t$ throughout this work, where it is obvious.

\subsection{Setup}
\begin{figure}
	\includegraphics[width=\textwidth]{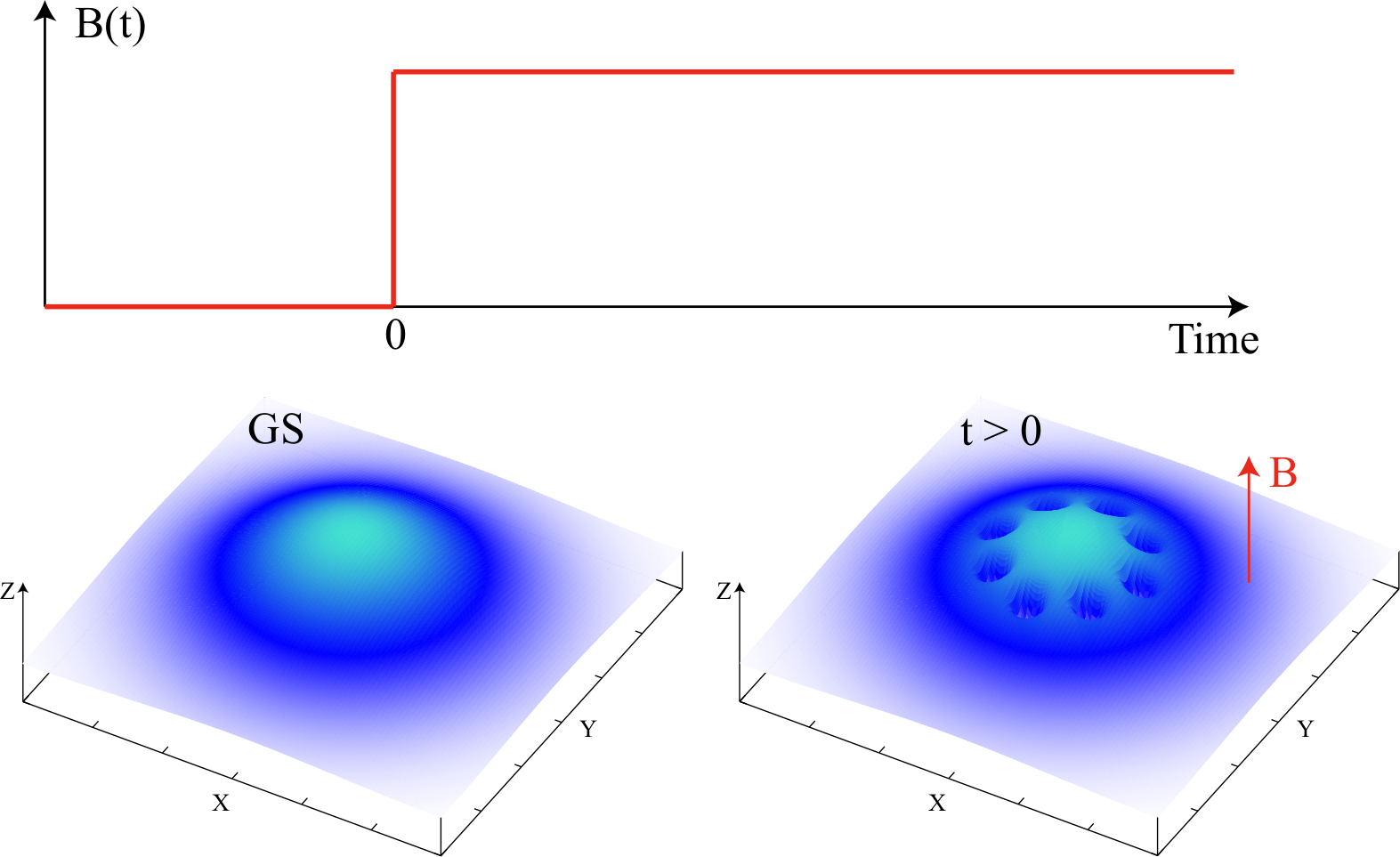}
	\caption{{\bf Sketch of our setup.} Two-dimensional ultracold bosonic particles are prepared in the ground state of an isotropic harmonic trap (label ``GS'') for time $t\leq0$. At $t>0$, an artificial magnetic field of strength $B(t)$ (top) is switched on suddenly. This quench triggers many-body dynamics of the state (label ``$t>0$'').}
	\label{fig:setup}
\end{figure}

To setup the time-dependent many-body Schrödinger equation (TDSE),
we use the Hamiltonian
\begin{equation}
\mathcal{H} = \int \mathrm{d} \mathbf{x} \: \hat{\Psi}^{\dagger}(\mathbf{x}) \left[ T(\mathbf{x}) + V(\mathbf{x}) \right] \hat{\Psi}(\mathbf{x}) \quad + \frac{1}{2} \int \mathrm{d} \mathbf{x} \mathrm{d} \mathbf{x}' \:  \hat{\Psi}^{\dagger}(\mathbf{x}) \hat{\Psi}^{\dagger}(\mathbf{x}') W(\mathbf{x},\mathbf{x}') \Psi(\mathbf{x}') \hat{\Psi}(\mathbf{x}). 
\label{eq:hamiltonian}
\end{equation}
Here, we work in atomic units ($\hbar=m=1$), the potential $V(\mathbf{x})$ [with $\mathbf{x}=(x,y)$] is chosen to be harmonic, $V(\mathbf{x})=\frac{1}{2}\mathbf{x}^2$, and we consider contact interactions $W(\mathbf{x},\mathbf{x}')=\lambda_0\delta(\mathbf{x}-\mathbf{x}')$. 
Formally, one cannot use a contact interaction in two spatial dimensions with a complete basis set~\cite{doganov:13,friedman:72} since the outcome would be that of the noninteracting bosons. In the present work, we employ a finite truncation of the many-body basis ($M=4$ orbitals) and aim to demonstrate that beyond-mean-field phenomena do emerge. 
The kinetic energy is augmented with an artificial gauge field $\mathbf{A}(\mathbf{x};t)$:
\begin{equation}
\hat{T}(\mathbf{x})= \frac{1}{2} \left(-i\nabla_{\mathbf{x}} - g \mathbf{A}(\mathbf{x};t) \right)^2.
\end{equation}
For simplicity, we consider the case of unit charge $g=1$ and a homogeneous magnetic field $\mathbf{B}$ in $z$-direction of strength $B(\mathbf{x};t)$:
\begin{equation}
\mathbf{B}(\mathbf{x};t) = B(t) \hat{e}_z. 
\end{equation}  
Here, $\hat{e}_z$ denotes the unit vector in $z$-direction.
In the following, we work in Landau gauge, 
\begin{equation}
\mathbf{A}(\mathbf{x};t)=B(t) \hat{e}_x,
\end{equation}
and consider a quench scenario in the following, i.e., 
\begin{equation}
B(t)= B \Theta(t).\label{eq:Boft}
\end{equation}
Here $\Theta(t)$ denotes the Heaviside step function, i.e., the magnetic field is suddenly turned on at $t>0$ after the system has been initialized, see Fig.~\ref{fig:setup} for a sketch.

In our investigation, we analyze the many-body dynamics by monitoring observables as a function of the effective magnetic field strength $B$ at $t>0$ after the quench.
\subsection{Method}
To solve the TDSE, 
\begin{equation}
\mathcal{H} \vert \Psi \rangle = i \partial_t \vert \Psi \rangle,
\end{equation}
we use the multiconfigurational time-dependent Hartree method for indistinguishable particles~\cite{alon.jcp:07,alon:08,streltsov:07,lode2:16,lode3:16,fasshauer:16,lin:20,lode:20}. Regarding our notation in Eq.~\eqref{eq:hamiltonian}, the MCTDH-X method implies that the field operators are represented by a sum of $M$ time-dependent single-particle states or orbitals: 
\begin{equation}
\hat{\Psi}(\mathbf{x})= \sum_{j=1}^{M} \hat{b}_j \phi_j(\mathbf{x};t).
\end{equation}
This corresponds to the following ansatz for the wavefunction:
\begin{equation}
\vert \Psi \rangle = \sum_{\vec{n}} C_{\vec{n}} \vert \vec{n} \rangle. \label{eq:ansatz}
\end{equation}
For bosons, the summation runs on all $\binom{N+M-1}{N}$ symmetric configurations $\vec{n}=(n_1,...,n_M)$ with fixed particle number $N=\sum_{i=1}^{M} n_i$. 
To derive the MCTDH-X equations, the ansatz in Eq.~\eqref{eq:ansatz} is plugged in a suitable variational principle~\cite{dirac:30,frenkel:34,mclachlan:64,kramer:81,haxton.pra2:15,kvaal:13}. The resulting equations are a two coupled sets -- a set of linear equations for the coefficients $\lbrace C_{\vec{n}}(t) \rbrace$ and a set of non-linear ones for the orbitals $\lbrace \phi_j(\mathbf{x};t); j=1,...,M\rbrace$, see Refs.~\cite{alon.jcp:07,alon:08,streltsov:07,lode2:16,lode3:16,fasshauer:16,lode:20} for details. In this work, we solve these equations with the MCTDH-X software~\cite{ultracold} hosted at \url{http://ultracold.org}.

\section{Quantities of Interest}\label{sec:QOI}
Here, we define the observables that we use to quantify the dynamics of $N$-boson systems: the one-body density, the eigenvalues of the reduced one-body density matrix (1BDM), the angular momentum, and the image entropy and its variance evaluated from simulated single-shot images.

\paragraph{\textbf{Density, One-body density matrix and Natural Occupations:}}

The 1BDM is a hermitian matrix defined as
\begin{equation}
\rho^{(1)}(\mathbf{x},\mathbf{x}') = \langle \Psi \vert\hat{\Psi}(\mathbf{x}) \hat{\Psi}^{\dagger}(\mathbf{x}')\vert  \Psi \rangle = \sum_{k,q} \rho_{kq} \phi^*_k(\mathbf{x};t) \phi_q(\mathbf{x};t).\label{eq:1BDM}
\end{equation}
Here, we used the matrix elements $\rho_{kq}=\langle \Psi \vert \hat{b}^\dagger_k \hat{b}_q \vert \Psi \rangle$ to represent the 1BDM using the orbitals corresponding to the creation and annihilation operators $\hat{b}^\dagger_k$ and $\hat{b}_q$, respectively.
The diagonal of the one-body density matrix is referred to as the density $\rho(\mathbf{x})$:
\begin{equation}
\rho(\mathbf{x}) = \rho^{(1)}(\mathbf{x},\mathbf{x}'=\mathbf{x}).
\end{equation}
The eigenvalues of the 1BDM, Eq.~\eqref{eq:1BDM}, can be obtained via a diagonalization that corresponds to a unitary transformation of the orbitals $\phi_j(\mathbf{x};t) $ to the natural orbitals $\phi_j^{(NO)}(\mathbf{x};t)$: 
\begin{equation}
\frac{\rho^{(1)}(\mathbf{x},\mathbf{x}')}{N} = \sum_j \lambda_j \phi_j^{(NO),*}(\mathbf{x};t) \phi_j^{(NO)}(\mathbf{x}';t).
\end{equation}
The eigenvalues $\lambda_j$ (natural occupations) determine the degree of condensation and fragmentation of the system. Bosons with a 1BDM with only a single contributing eigenvalue $\lambda_1$ are condensed~\cite{penrose:56} and bosons with a 1BDM with multiple macroscopic eigenvalues contributing  $\lambda_1\mathcal{O}(N);\lambda_2\mathcal{O}(N);...$ are fragmented~\cite{nozieres:82,spekkens:99}.

\paragraph{\textbf{Angular momentum:}}
The angular momentum operator in $\hat{e}_z$-direction for a two-dimensional system is defined as 
\begin{equation}
\hat{L}_z = \hat{e}_z (\hat{x} \times \hat{p}) = - i \left( \hat{x} \hat{\partial}_y - \hat{y}\hat{\partial}_x \right).\label{eq:AM}
\end{equation}
Bosonic quantum systems with angular momentum are rich in physics: they feature condensed vortices~\cite{gross:61,pitaevskii:61,wells:15}, phantom vortices~\cite{sakmann:16,weiner:17}, spatially partitioned many-body vortices~\cite{beinke:15,klaiman:16c}, and fragmentation~\cite{tsatsos:10,tsatsos:15,beinke:15,klaiman:16c,sakmann:16,weiner:17}.

\paragraph{\textbf{Single shots, image entropy and its variance:}}
To assess the observability of the emergent physics in experimental setups with ultracold atoms, we simulate the detection of our numerical model wavefunctions in absorption or single-shot images~\cite{sakmann:16,gajda:16,lode:17}. A set of $N_s$ single shots,
\begin{equation}
\mathcal{\mathbf{S}}^j=(\mathbf{s}_1^j,...,\mathbf{s}^j_N);\qquad j=1,...,N_s,
\end{equation}
is nothing but $N_s$ random samples that are $N$-variate and distributed according to the $N$-particle probability given by $\vert \Psi \vert^2$,
\begin{equation}
P(\mathbf{x}_1,...,\mathbf{x}_N) = \vert \Psi(\mathbf{x}_1,...,\mathbf{x}_N) \vert^2.
\end{equation}
To generate images from these single shots, we convolute with an point spread function. Typical choices include Gaussian (see \cite{sakmann:16,lode:17,chatterjee:18,chatterjee:20}) or even quantum point spread functions~\cite{pyzh:19}. Here, for simplicity, we consider the idealized case of a $\delta$-shaped point spread function to obtain our single-shot images:
\begin{equation}
\mathcal{S}^j(\mathbf{x})= \sum_{i=1}^N \delta(\mathbf{x}-\mathbf{x}^j_i).
\end{equation}
We will consider the image entropy $\zeta$ of single-shot images of the state $\vert \Psi \rangle$:
\begin{equation}
\zeta = - \frac{1}{N_s} \sum_{j=1}^{N_s} \zeta^j; \qquad \zeta^j= \int d\mathbf{x} \mathcal{S}^j(\mathbf{x})  \ln \mathcal{S}^j(\mathbf{x}) . \label{eq:entropy} 
\end{equation}
In the limit of large $N_s$, the image entropy $\zeta$ is equivalent to the density-entropy studied, for instance, in Ref.~\cite{roy:18}. Variances of observables serve as a precursor of quantum fluctuations and correlations in many-body systems~\cite{klaiman:15,klaiman:16b,klaiman:18,alon:18b,alon:19}; we are thus motivated to also analyze the variance $\sigma_\zeta$ of the image entropy $\zeta$:
\begin{equation}
\sigma_\zeta = \frac{1}{N_s} \sum_{i=1}^{N_s} \left[\zeta - \zeta_j \right]^2. \label{eq:var_entropy}
\end{equation}

\section{Results}\label{sec:RES}
We now carve out the connection between artificial gauge fields and many-body correlations. For this purpose we focus on the dynamics of a model system of $N=100$ two-dimensional ultracold bosonic atoms with an interaction strength of $\lambda_0=0.01$ [cf. Eqs.~\eqref{eq:hamiltonian}--\eqref{eq:Boft} for $t\leq0$]. The system is initialized in its ground state and its dynamics ($t>0$) are then triggered by suddenly turning on an AMF of strength $B$ [Eq.~\eqref{eq:Boft}]. In what follows, we aim at an understanding of how the strength of the AMF affects the emergent dynamical behavior. For this purpose, we solved the time-dependent many-body Schrödinger equation with MCTDH-X using $M=4$ orbitals [$176581$ configurations in the state in Eq.~\eqref{eq:ansatz}] and $128\times128$ DVR functions to represent each of the orbitals $\lbrace \phi_j(\mathbf{x};t)\rbrace$.

We open the exposition of our findings with the density $\rho(\mathbf{x})$ and its decomposition into natural orbitals $\phi_j^{(NO)}$.
\begin{figure}
	\includegraphics[width=\textwidth]{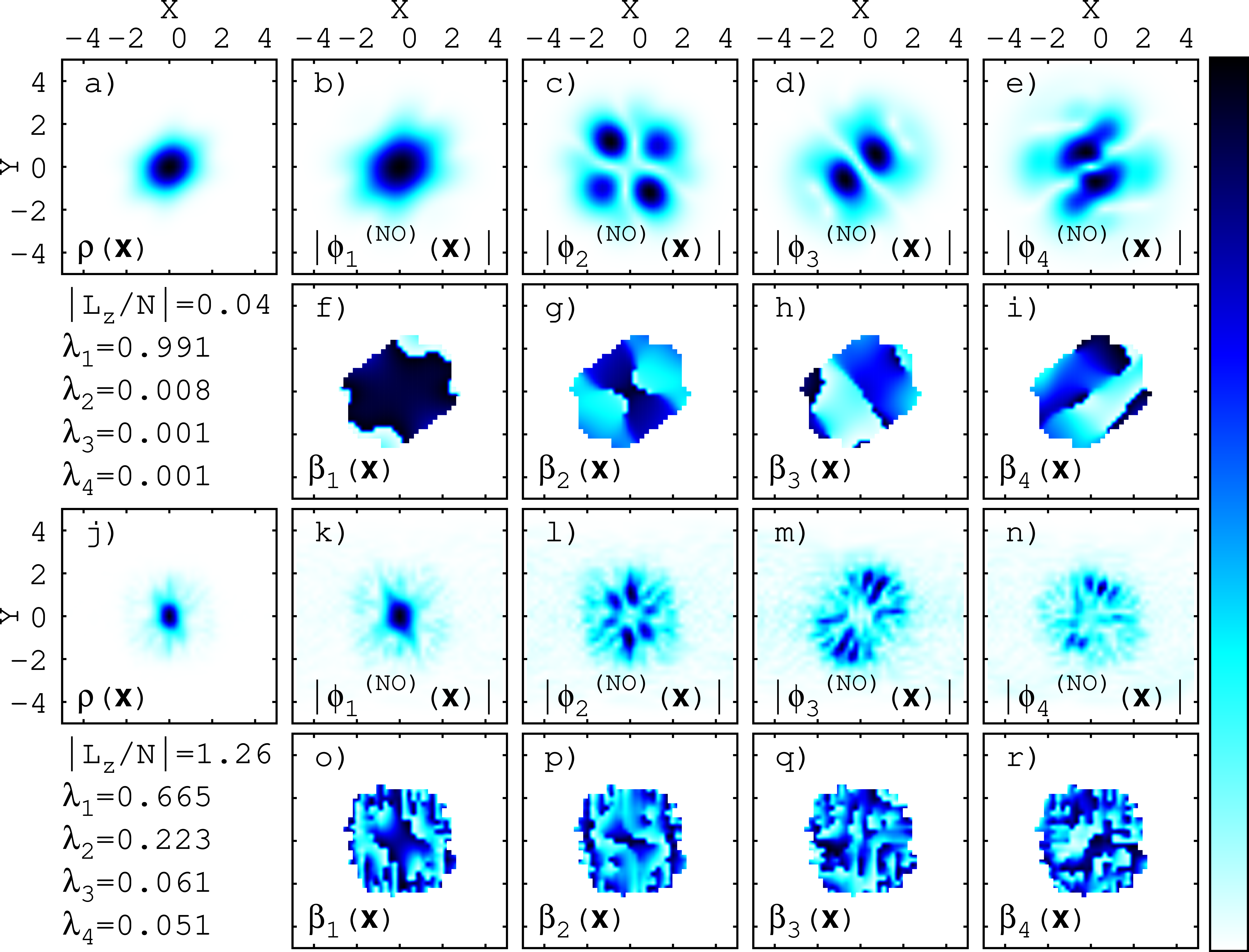}
	\caption{{\bf One-body density, natural orbitals, and natural orbital phases} for two distinct AMF strengths: $B=1.0$ (weak) in a)--i) and $B=6.5$ (strong) in j)--r). The figure shows the quantities at time $t=50.0$. 
		For guidance, the angular momentum per particle and the natural occupations are listed below a) and j) for weak and strong AMFs, respectively. The phase plots [$\beta_j(\mathbf{x})$ in f)--i) and o)--r)] are restricted to areas where $\rho(\mathbf{x})>0.01$; see text for further discussion.}
	\label{fig:RHO}
\end{figure}

Besides a slight deformation of the density and the orbitals, little effects are seen in Fig.~\ref{fig:RHO}a)--e) for a weak AMF, $B=1.0$.  The phases $\beta_{1/2}(\mathbf{x})$, f)--g), hint there are no phantom vortices. The phases $\beta_{3/4}(\mathbf{x})$ feature topological defects aligned with zeros in $\phi_{3/4}(\mathbf{x})$, but these orbitals are occupied only by $0.1$ particles.

For a comparatively strong AMF, $B=6.5$, in contrast, vortices at the edges of the density (so-called ``ghost vortices''~\cite{tsubota:02}) and phantom vortices~\cite{weiner:17} in the orbitals emerge in Fig.~\ref{fig:RHO}j)--r): zeros of the orbital densities are accompanied by topological defects in their phase [compare Fig.~\ref{fig:RHO}, panels k) and o), l) and p), m) and q), n) and r), respectively]. 

These features of the density, orbitals, and their phases are hallmarks of the angular momentum that is deposited in time by the action of a sufficiently strong AMF: for increasing AMF strength $B$, the expectation value $L_z=\langle \Psi \vert \hat{L}_z \vert \Psi \rangle$ of the angular momentum operator $\hat{L}_z$ [Eq.~\eqref{eq:AM}] increases. For instance, we find $L_z/N = 0.04$ at time $t=50.0$ and $B=1.0$ in Fig.~\ref{fig:RHO}a)--i) and $L_z/N = 1.26$ at time $t=50.0$ for $B=6.5$ in Fig.~\ref{fig:RHO}j)--r)].

\begin{figure}
	\includegraphics[width=\textwidth]{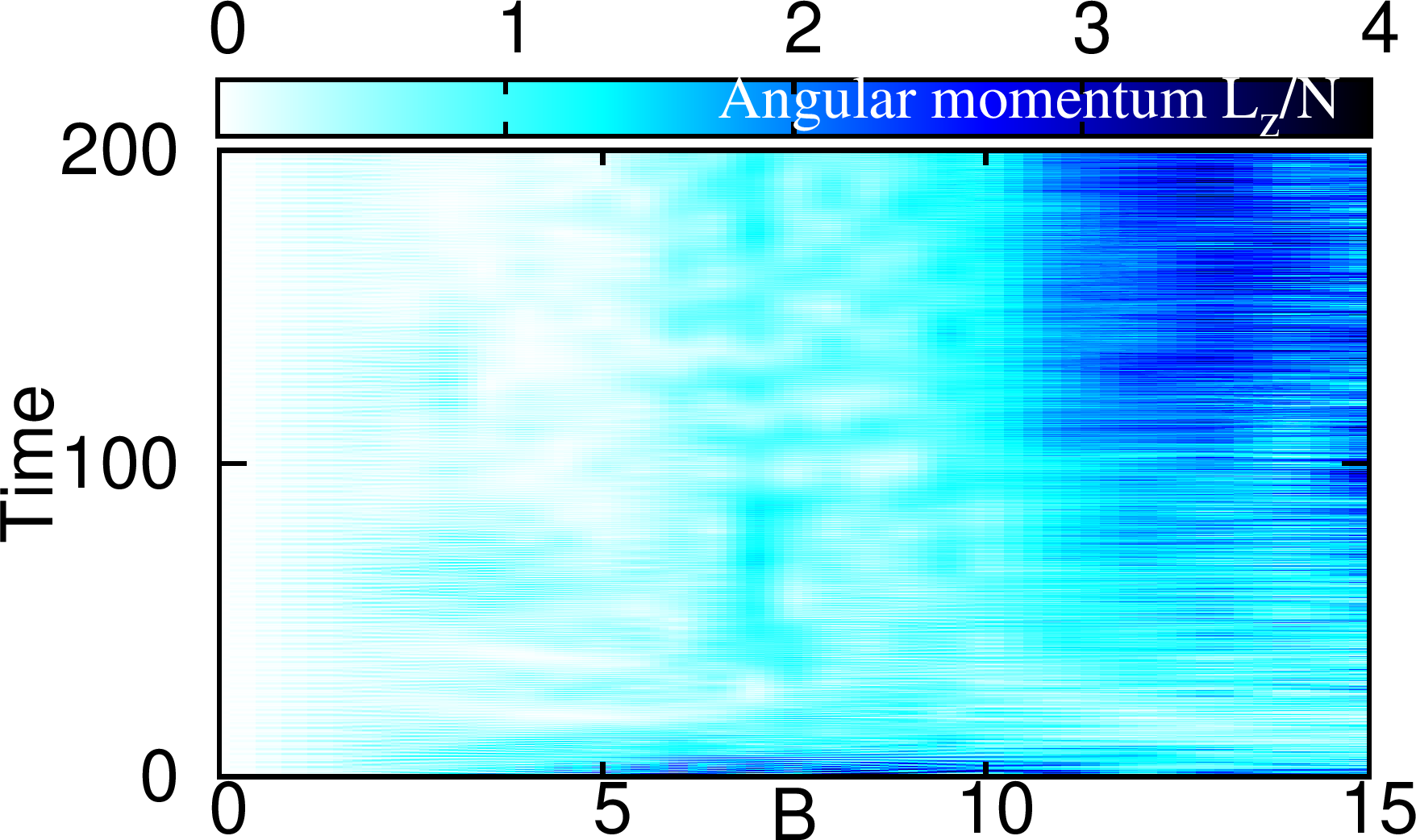}
	\caption{{\bf Angular momentum as a function of propagation time and strength of AMF.} The expectation value $L_z/N=\langle \Psi(t) \vert \hat{L}_z \vert \Psi(t) \rangle/N$ [cf. Eq.~\eqref{eq:AM}] is shown as a function of time $t$ and AMF strength $B$. Interestingly, there is a drastic short-term increase following a subsequent equilibration of $L_z/N$ for $5\lesssim B \lesssim 12$. See text for discussion.}
	\label{fig:LZ}
\end{figure}

\begin{figure}
\includegraphics[width=\textwidth]{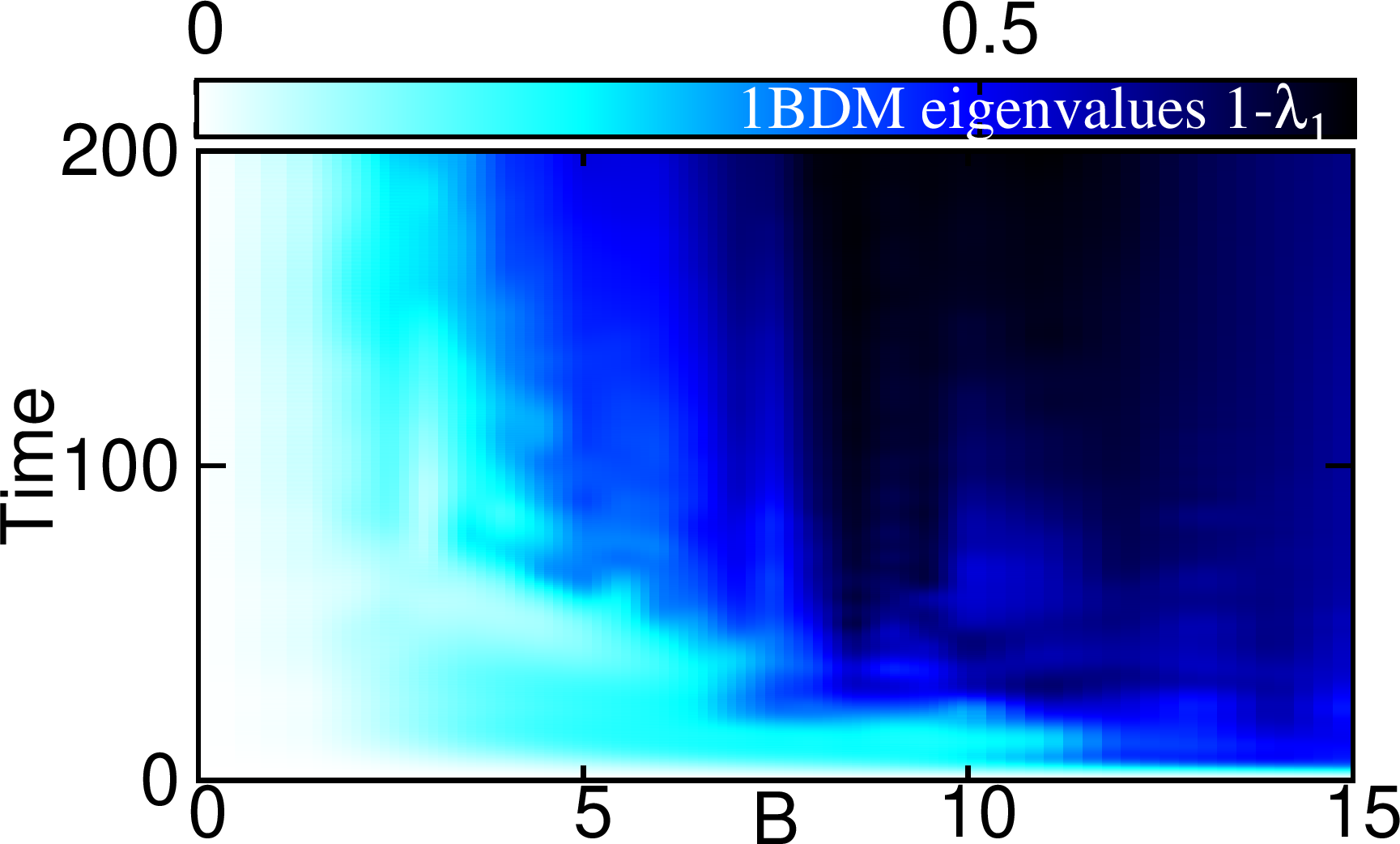}
\caption{{\bf Eigenvalues of the reduced one-body density matrix.} The dynamical emergence of multiple significant eigenvalues, i.e., the fragmentation of the many-body state, is quantified here via the time-dependent depletion $1-\lambda_1=\sum_{k=2}^{M} \lambda_k$ and is in sync with the dynamics of angular momentum (cf. Fig.~\ref{fig:LZ}).}
\label{fig:NO}
\end{figure}

To quantify the dynamics of angular momentum triggered by quenches of the AMF a bit better, we plot $L_z/N$ for our system as a function of evolution time and as a function of the strength of the AMF in Fig.~\ref{fig:LZ}.
We find from Fig.~\ref{fig:LZ} that a threshold AMF strength of about $B\gtrsim 6 $ is required to generate states with non-zero angular momentum at long evolution times (here, $t=200$). Furthermore, the average angular momentum content increases as the strength of the AMF does. In Refs.~\cite{mueller:06,dagnino:09,tsatsos:10,tsatsos:15,beinke:15,HLRS:15,cremon:15,sakmann:16,weiner:17,HLRS:17,eriksson:19}, an intricate connection of angular momentum content and the presence of correlations or the fragmentation of many-boson states has been pointed out. This motivates us to analyze the time-evolution of the eigenvalues of the reduced one-body density matrix as a precursor of correlations and the departure of the analyzed state from a mean-field description; we, thus, underpin the limitations of a mean-field description, see Fig.~\ref{fig:NO} for a plot of $\lambda_j$ as a function of time and strength $B$ of the AMF. 
Our present findings for a quenched AMF are in line with results obtained  with a time-dependent and slow transfer of angular momentum via a rotating asymmetry of the harmonic trapping potential~\cite{sakmann:16,weiner:17}: the dynamical departure from a single-eigenvalue 1BDM to a correlated many-body state is in-sync with the dynamical acquisition of an angular momentum [compare Fig.~\ref{fig:NO} and Fig.~\ref{fig:LZ}]. 

\begin{figure}
	\includegraphics[width=\textwidth]{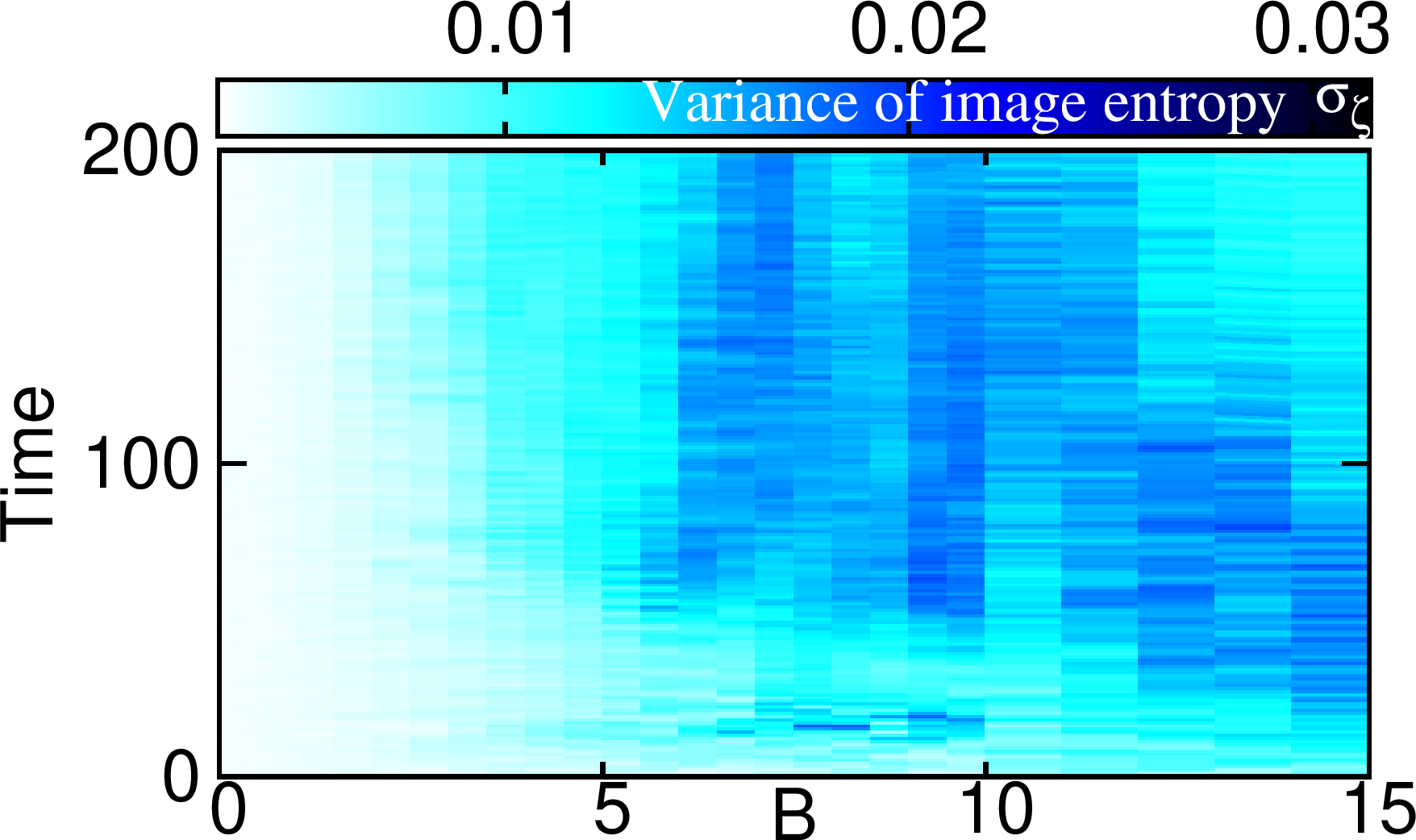}
	\caption{{\bf Variance of the image entropy as a function of propagation time and strength of the AMF.} The variance of image entropy $\sigma_\zeta$ tracks the behavior of the natural occupations and the angular momentum in, respectively, Figs.~\ref{fig:NO}, and ~\ref{fig:LZ} closely, see text for discussion.}
	\label{fig:VARZETA}
\end{figure}

We now turn to the question of the possibility of an experimental detection of the emergent behavior of angular momentum and the eigenvalues of the one-body density matrix. For this purpose we simulated $N_s=1000$ single-shot images for all the many-body wavefunctions $\vert \Psi (t) \rangle$ for every time in $t\equiv k \cdot dt \in [0,200]$ in steps of $dt=1.0$. From this dataset of single-shot images, we computed the image entropy (see Appendix~\ref{sec:ZETA}) and its variance $\sigma_\zeta$ [Eq.~\eqref{eq:var_entropy}]; see Fig.~\ref{fig:VARZETA}.
The variance $\sigma_\zeta$ follows the behavior of angular momentum and natural occupations in Figs.~\ref{fig:NO} and ~\ref{fig:LZ}, respectively. This clearly demonstrates that an experimental detection of the presence of correlations and angular momentum is feasible.

\section{Conclusions and outlook}\label{sec:END}
We analyzed the dynamics of interacting two-dimensional ultracold bosonic particles triggered by a quench of an artificial gauge field.
We used the multiconfigurational time-dependent Hartree method for indistinguishable particles software (\url{http://ultracold.org}) to solve the many-body Schrödinger equation from first principles.
Our exploratory investigation demonstrates that many-body correlations emerge due to the quench if the artificial magnetic field is sufficiently strong. We have portrayed how these correlations show in the natural occupations, the expectation value of the angular momentum operator, and in the orbitals and their phases as phantom vortices. 
Using simulations of single-shot images, we demonstrate that the correlations can be detected via the variance of the entropy of the images.

Our work highlights the importance of deploying modern computational and theoretical many-body approaches like the MCTDH-X to systems with artificial gauge fields as well as the necessity to consider not only the wavefunctions of ultracold atoms themselves, but also their detection. 

Our results complement the recent findings, that the variances of observables are sensitive probes of correlations in the state of ultracold atomic systems~\cite{klaiman:15,klaiman:16b,HLRS:17,lode:17,alon:18,klaiman:18,alon:18b,alon:19,lode:20}.

Straightforward continuations of this investigation include the deployment of the developed analysis and computational tools to other many-body systems. As examples of interest, we name here the variance of the image entropy in ultracold dipolar atoms as discussed in Refs.~\cite{chatterjee:18,chatterjee:19,chatterjee:20} and an exploration of the competition of long-ranged dipolar interactions and artificial magnetic fields for two-dimensional ultracold atoms.
Another direction of physical interest is the dynamics of two-dimensional bosonic Josephson junctions~\cite{bhowmik:20} subject to gauge fields and the resulting tunneling of (phantom) vortices or the emergence of quantum turbulence~\cite{tsatsos:16} via entropy production~\cite{madeira:20}, which, in turn, as we have shown above, could result from the presence of artificial gauge fields.

\section{Complementary results on image entropy}\label{sec:ZETA}
In this appendix we show complementary results on the image entropy $\zeta$ as a function of AMF strength and evolution time. 

\begin{figure}
	\includegraphics[width=\textwidth]{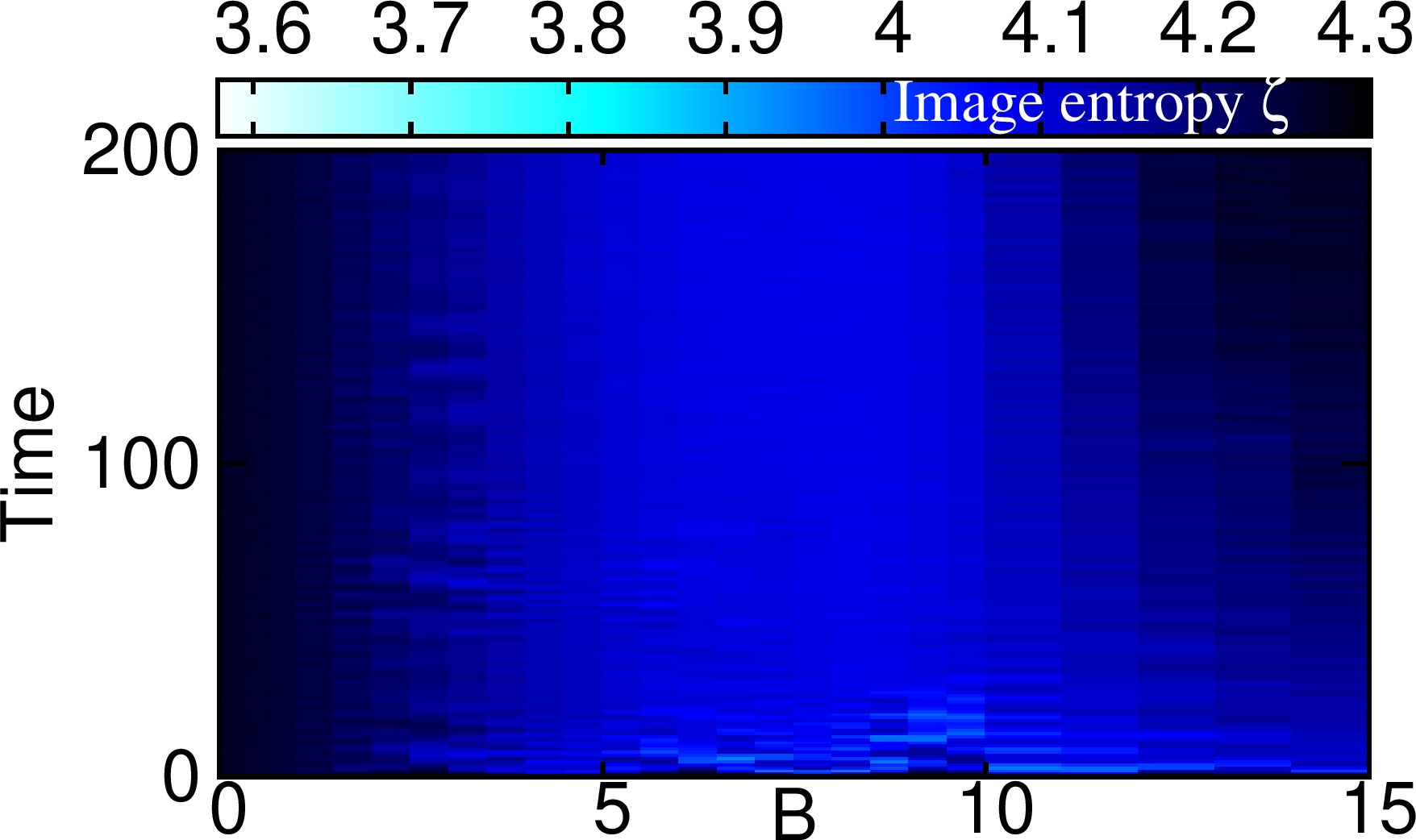}
	\caption{{\bf Image entropy as a function of propagation time and strength of AMF.} The image entropy $\zeta$ [Eq.~\eqref{eq:entropy}] shows little overall variation as a function of AMF strength $B$ and evolution time.}
	\label{fig:ZETA}
\end{figure}
The entropy $\zeta$ [Eq.~\eqref{eq:entropy}] in
Fig.~\ref{fig:ZETA} 
is lowest, where its variance (Fig.~\ref{fig:VARZETA}) is changing the most as a function of AMF strength $B$ and time. 
The exploration of a possible fundamental connection of the gradient of the image entropy and the variance $\sigma_\zeta$ goes beyond the scope of our present investigation.

\authorcontributions{A.U.J.L. performed computations, A.U.J.L. and S.D. implemented numerical algorithms, and A.U.J.L., S.D., and C.L. designed the study and wrote the manuscript.}
\funding{This work and A.U.J.L. has been supported by the Austrian Science Foundation (FWF) under grant P-32033-N32. C.L. acknowledges funding by the FWF under grant M-2653 and by The Wiener Wissenschafts und Technologie Fonds (WWTF) project No. MA16-066. Computation time on the Hawk cluster at the HLRS Stuttgart is acknowledged. Support by the German Research Foundation (DFG) and the state of Baden-W\"urttemberg via the bwHPC grants no INST 40/467-1 FUGG (JUSTUS cluster), INST 39/963-1 FUGG (bwForCluster NEMO), and INST 37/935-1 FUGG (bwForCluster BinAC) is gratefully acknowledged.}
\conflictsofinterest{The authors declare no conflict of interest.}
\acknowledgments{Insightful discussions with Saurabh Basu and Ofir E. Alon are gratefully acknowledged}
\section*{References}
\bibliography{AG}
\end{document}